\documentclass[fleqn,usenatbib,useAMS,letters]{mnras}
\usepackage{xcolor}
\usepackage{amsmath}    
\usepackage{multicol}        
\usepackage{bm}         
\usepackage{pdflscape}  
\usepackage{graphicx}

\usepackage[T1]{fontenc}
\usepackage{ae,aecompl}

\usepackage{newtxtext,newtxmath}

\newcommand{\kms}{km\,s$^{-1}$}

\usepackage[normalem]{ulem}
\defcitealias{Goudfrooij18a}{Paper~I}
\newcommand{\lea}{{\>\rlap{\raise2pt\hbox{$<$}}\lower3pt\hbox{$\sim$} \>}}
\newcommand{\gea}{{\>\rlap{\raise2pt\hbox{$>$}}\lower3pt\hbox{$\sim$} \>}}

\newcommand{\FUVmV}{${\it FUV}\!-\!V$}

\newcommand{\aFe}{[$\alpha$/Fe]}

\newcommand{\UV}{{\rm F275W}}
\newcommand{\WashC}{{\rm F390W}}

\newcommand{\g}{{\rm F475W}}

\newcommand{\z}{{\rm F850LP}}
\newcommand{\HST}{{\it HST}}
\newcommand{\mydel}{$\Delta$}
\newcommand{\UVminC}{${\rm F275W}-{\rm F390W}$}
\newcommand{\gminz}{${\rm F475W}-{\rm F850LP}$}
\newcommand{\SnMR}{$S_{\rm N,\,MR}$}

\newcommand{\pgscriptsize}{\fontsize{7.5}{8.5}\selectfont}

\title[The UV upturn vs.\ multiple stellar populations in GCs]{A colour-colour fingerprint links the UV upturn in early-type galaxies to second-generation stars from dissolved globular clusters\thanks{Based
  on observations made with the NASA/ESA Hubble Space Telescope, obtained at the
  Space Telescope Science Institute, which is operated by the Association of
  Universities for Research in Astronomy, Inc., under NASA contract
  NAS5-26555.}}

\author[P. Goudfrooij et al.]{Paul Goudfrooij,$^{1}$%
\thanks{Contact e-mail: \href{mailto:goudfroo@stsci.edu}{goudfroo@stsci.edu}}
Andrea Bellini,$^{1}$ Thomas M. Brown,$^{1}$
and Thomas H. Puzia$^{2}$ 
\\
$^{1}$Space Telescope Science Institute, 3700 San Martin Drive, Baltimore, MD 21218, USA \\
$^{2}$Institute of Astrophysics, Pontificia Universidad
  Cat\'olica de Chile, Av.\ Vicu\~{n}a Mackenna 4860, Macul  
  7820436, Santiago, Chile
  }
\date{Accepted 2026 June 5. Submitted 2026 June 3; in original form
  2026 May 5}

\pubyear{2026}

\begin{document}
\label{firstpage}
\pagerange{\pageref{firstpage}--\pageref{lastpage}}
\maketitle

\begin{abstract}
We address two mass-dependent properties among early-type galaxies
(ETGs): (1) abundance ratios [N/Fe] and [Na/Fe], and (2) the centrally
concentrated ``UV upturn'' at far-UV (FUV) wavelengths, which is
likely produced by extreme horizontal branch stars with supersolar helium
abundances. Using new \emph{HST/WFC3} observations of one 
FUV-weak and one FUV-bright ETG, we probe the ``MP scenario'' by
Goudfrooij who posited that the UV upturn and the mass-dependent
abundance variations of N and Na within and among ETGs are
physically connected and produced by dissolution of metal-rich
globular clusters, which represent the only galactic environment where
mass-dependent enrichment of He, N, and Na is known to occur (i.e.,
second-generation stars of the ``multiple stellar populations'' (MPs)
phenomenon). We show that passbands F275W and F390W are uniquely
sensitive to correlated changes in $Y$ and [N/Fe] in integrated-light
photometry when combined with archival data in F475W and F850LP.
While \gminz\ is found to decrease with increasing radius in both
galaxies, consistent with known metallicity gradients, \UVminC\ 
increases with increasing radius, as expected if the UV upturn is caused by
second-generation stars with supersolar $Y$ and [N/Fe]. Furthermore, the
radial gradient in \UVminC\ and the implied fractions of He-- and
N-enhanced stars are found to be significantly larger in the
FUV-bright ETG than in the FUV-weak one, consistent with the
predictions of the MP scenario.    
\end{abstract}



\section{Introduction} \label{s:intro}
Historically, the smooth appearance, lack of significant interstellar matter, and red color of early-type galaxies (hereafter ETGs), was generally interpreted as them consisting of old, metal-rich stellar populations in which all star formation has ended long ago. ETGs have long been known to exhibit super-solar $\alpha$-element abundances \citep[e.g.,][]{Worthey+92, Trager+00}, with the observed correlation of \aFe\ with galaxy mass being interpreted as an inverse relation between mass and star formation time scale \citep[e.g.,][]{Thomas+05}. However, more recently, several lines of evidence have revealed that the chemical abundance ratios of massive ETGs are 
more complex than the expectation for stars formed in short bursts with ``simple'' $\alpha$ enhancement. 

One example of this was introduced through modeling of individual element abundances in integrated-light spectroscopy, finding that [N/Fe] and [Na/Fe] also increase with galaxy mass \citep*[e.g.,][]{Schiavon07, Conroy+14}. Furthermore, many ETGs are much more luminous in the far-ultraviolet (FUV; $\lambda \lesssim$ 200 nm) than the expectation for an old, metal-rich stellar population. This feature is now widely known as the ``UV upturn'' \citep[e.g.][]{CodeWelch79, OConnell99}. Its strength is measured by the \FUVmV\ colour, which is inversely correlated with the velocity dispersion (i.e., mass) and metallicity of the host galaxy \citep{Burstein+88,Dorman+95,Jeong+12,Goudfrooij18a}. Several FUV imaging and spectroscopy studies have established that the bulk of the FUV flux in ETGs originates from very hot ($T_{\rm eff} \ga$\, 20,000 K) helium-burning extreme horizontal branch (EHB) stars and their progeny \citep[e.g.,][]{OConnell99,Brown00}. However, for metal-rich populations such as those in massive ETGs, EHB stars can only attain such high temperatures if their helium abundance is significantly super-solar \citep[e.g.,][]{Yi+05}.  

A key property of the UV upturn in ETGs in this context is that the FUV emission is significantly more centrally concentrated than the optical light \citep{Ohl+98, Carter+11}, thus suggesting that the helium enhancement has mainly accumulated in the central regions of ETGs. As such, there is now strong evidence that the super-solar \aFe\ in ETGs is accompanied by super-solar [N/Fe], [Na/Fe], and helium abundances in their central regions. 

The only galactic environment where both $\alpha$ enhancement \citep*{Puzia+06} and enhancement of He, N, and Na is \emph{known} to exist is in massive globular clusters (GCs). Recent photometric and spectroscopic observations of GCs in the Local Group have established that GCs exhibit significant internal spreads of light-element abundances, with the more massive GCs also hosting stars significantly enriched in He \citep[e.g.,][]{Piotto+07, Milone+18}. Well-known features of these spreads are the Na-O and N-O anti-correlations among stars within GCs, whose extent scales with GC mass \citep[e.g.,][]{Carretta+10}.

This phenomenon is commonly referred to as ``multiple stellar populations'' (MSPs) within GCs, consisting of two main stellar populations: the first one, often referred to as the ``first generation'' (FG), consists of stars that show chemical abundance patterns consistent with ``normal'' $\alpha$-enhanced populations, while the ``second-generation'' (SG) population shows enhanced He, N, Na, and Al, along with depletions of O, C, and Mg. Even though no single scenario for the origin of these features is free from shortcomings \citep{Renzini+15}, they are thought to result from GC self-enrichment by means of retention of material from ``polluter'' FG stars that undergo proton capture reactions at high temperatures at late evolutionary stages, facilitating the CNO, Ne-Na, and O-N cycles \citep{BastianLardo18, Gratton+19}. The fraction of SG stars found within Galactic GCs increases strongly with increasing GC mass, ranging from $\la$\,30\% at $10^{4.5}\;M_{\odot}$ up to $\sim$\,90\% at $10^{6.5}\;M_{\odot}$ \citep{Milone+17}. Similarly, the spread in He abundance ($Y$) found in Galactic GCs is strongly correlated with GC mass \citep{Milone+18}, similar to the case of [N/Fe] and [Na/Fe] spreads mentioned above.  

With the above in mind, \citet{Goudfrooij18a} proposed a scenario in which the UV upturn in ETGs is due to SG stars that were formed in young massive star clusters and subsequently dispersed into the field population due to dynamical dissolution of these clusters in the strong tidal field of the central regions of the host galaxy \citep*[see also][]{Chantereau+18}. The evidence underlying this scenario was the finding of a strong correlation between UV upturn strength and the specific frequency of metal-rich GCs in ETGs (hereafter \SnMR). The latter is defined as the number of GCs in the metal-rich peak of the well-known bimodal optical colour distribution of GCs in ETGs 
\citep[e.g.,][]{Peng+08}, normalized by the galaxy luminosity for which we use the Sloan $z$ band. \SnMR\ is thought to reflect the fraction of stars formed in (metal-rich) star clusters at the massive end of the initial cluster mass function, many of which survive a Hubble time of dissolution processes \citep[e.g.,][]{McLaughlin99}. That such dissolution processes are at work is supported by steep positive gradients of radial GC size distributions in ETGs \citep{Puzia14}. According to \citet{Goudfrooij18a}, the correlation between \SnMR\ and \FUVmV\ then arises through higher fractions of stars having originated from the most massive star clusters (which have the highest fractions of SG stars) in the more massive ETGs, especially in their central regions where the tidal field is strongest.  

In this Letter we report early results from a {\it Hubble Space Telescope (HST)} program for which the aim is to address the question: can we find observational evidence for the presence of significant numbers of SG stars in the central regions of ETGs, and if so, does the overabundance of both He and N scale with \SnMR\ and \FUVmV\ in a manner consistent with the MP scenario?

\section{Method} \label{s:method}
To establish an appropriate photometric method to detect radial gradients of overabundances of He and N in ETGs, we construct integrated-light spectral energy distributions (SEDs) of stellar populations from a large grid of synthetic spectra and chemical compositions typical of first- and second-generation stars found in massive, high-metallicity GCs in the Milky Way. In this context, we use the suite of codes ATLAS12 and SYNTHE developed by R.\ L.\ Kurucz and F.\ Castelli \citep*[see][]{Castelli05, Kurucz05, Sbordone+07} which allows one to use arbitrary chemical compositions. To do so, we follow the methodology described in the Appendix of \citet{goukru13} with the following exceptions due to findings that emerged after that study.

To simulate GCs in the central regions of ETGs, we choose [$Z$/H] = $-$0.2 which is consistent with the mass-weighted average [$Z$/H] of massive ETGs within their effective radius as found by extensive spectroscopic studies \citep{McDermid+15,Martin-Navarro+21}. To represent abundances of He, C, N, O, Na, Mg, and Al in FG and SG stars in metal-rich GCs, we average values found in two massive metal-rich Galactic GCs located in the Milky Way bulge: NGC 6388 and NGC 6441, both of which contain blue horizontal branches that extend blueward of the RR Lyrae instability strip, i.e., the type of stars that are thought to produce the UV upturn in ETGs.  For FG stars, we select the primordial He abundance ($Y$ = 0.235 + 1.5 $Z$) and choose [C/Fe] = 0.06, [N/Fe] = 0.20, [O/Fe] = 0.40, [Na/Fe] = 0.00, [Mg/Fe] = 0.25, and [Al/Fe] = 0.00 from \citet{Cannon+98}, \citet{Gratton+06}, \citet{Carretta+09}, and \citet{carbra18}.  For SG stars, we use the following \emph{mean}\footnote{Each [X/Fe] ratio as well as $Y$ encompasses a \emph{range} within GCs \citep{Milone+18}. We model this by using mean values for SG stars.} abundance differences relative to FG stars: \mydel$Y$ = 0.05, 
\mydel[C/Fe] =  $-$0.55, \mydel[N/Fe] = 1.00, \mydel[O/Fe] = $-$0.45, \mydel[Na/Fe] = 0.35\footnote{We use \mydel[Na/Fe] = 0.432 \mydel[N/Fe] from \citet{Yong+05}.}, \mydel[Mg/Fe] = $-$0.10, and \mydel[Al/Fe] = 0.75 \citep[see][and references therein]{Milone+18}. We make these choices because actual abundance ratios are available for these clusters in the literature. Note, however, that there is evidence for even higher helium abundances in at least some massive GCs in the central regions of the massive ETG M87 \citep{Kaviraj+07,Bellini+15}. As such, the differences in abundance ratios between FG and SG stars in massive ETGs could be stronger than those assumed here.  

For the creation of stellar model atmospheres, we use the isochrones of \citet{Bertelli+08} for an age of 12 Gyr. These isochrones include the horizontal branch, and they were calculated for various He abundances. The isochrones are used for both the FG and SG stars, since stellar evolution is not affected by these abundance variations if the sum of C, N, and O abundances is kept constant \citep[see, e.g.,][]{Sbordone+11}, which is the case for our choices. For each isochrone, we sample the full $\log(L) - \log(T_{\rm eff})$ parameter space in a semi-uniform manner using $\sim$\,25 ($L$, $T_{\rm eff}$) combinations, making sure that regions in parameter space where changes are relatively rapid are properly sampled.  

After the individual stellar spectra are calculated for a given population, they are summed together by weighting them by their initial stellar masses ${\cal{M}}_i$ and differences between them ($\Delta\,{\cal{M}}_i$) according to a \citet{Kroupa01} initial mass function, thus producing integrated-light spectra for the full stellar population.

We also calculated several other model spectra for comparison purposes: one such set of spectra was created by varying one element at a time so as to evaluate their respective effects on photometric colours (see Sect.~\ref{s:results} below). To compare the effects of light-element abundance variations as a function of galactocentric radius with those of age and metallicity gradients, we also created a FG model spectrum for [$Z$/H] = $-$0.5 and one for a younger age (8.5 Gyr).

\begin{figure*}
\includegraphics[width=10.5cm]{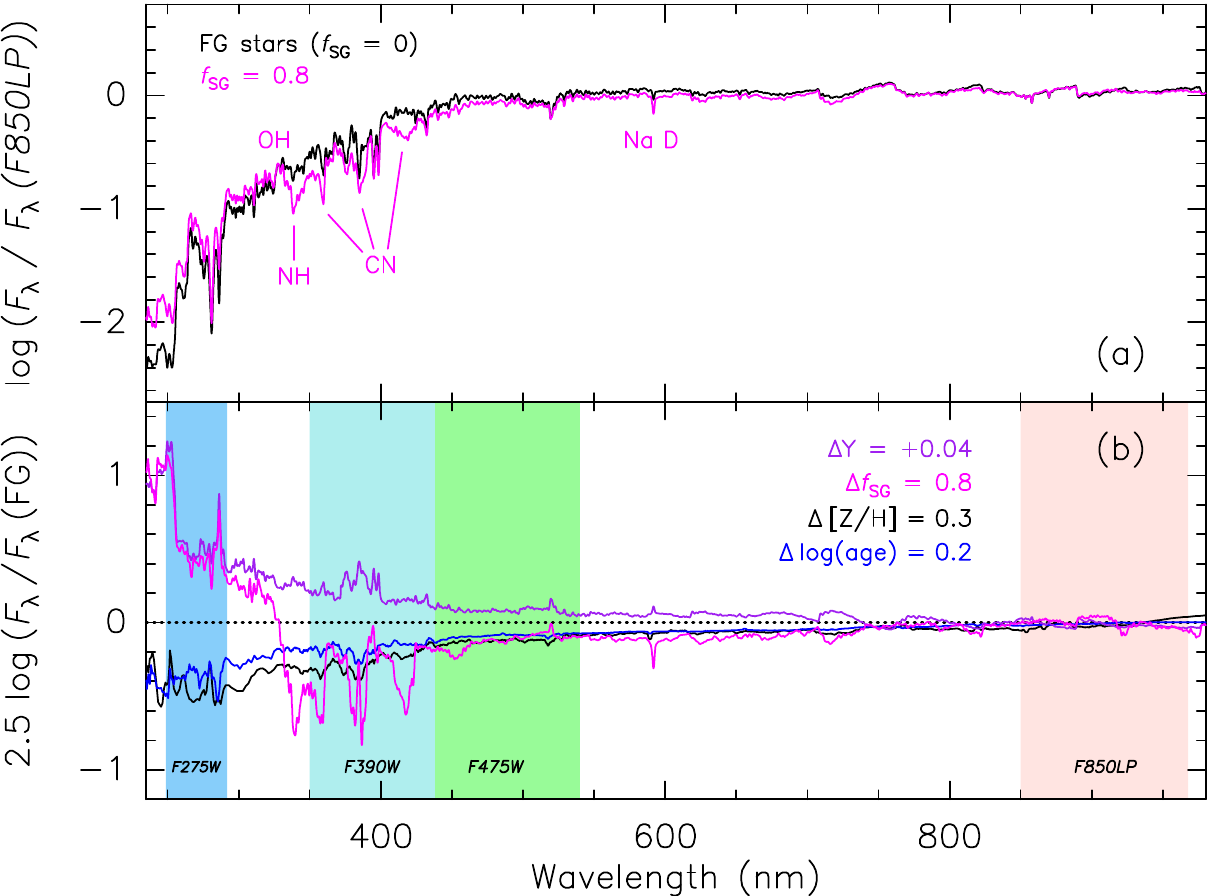}
\caption{\emph{Panel (a)}: model SSP spectra for an age of 12 Gyr and [$Z$/H] = $-$0.2 in the 200\,--\,1000 nm range. The spectra are rendered at $R$ = 500, redshifted to $V_{\rm rad} = 1280$ \kms, appropriate for the Virgo cluster, and normalized at the F850LP passband of {\it HST/ACS}. The black line represents the FG spectrum ($f_{\rm SG}$ = 0), while the magenta line represents a population typically seen in massive, metal-rich GCs (with a SG fraction $f_{\rm SG}$ = 0.8). Relevant spectral features are highlighted. \emph{Panel (b)}: ratio of spectra relative to the FG spectrum, normalized in the F850LP passband and expressed in magnitudes. Relevant \emph{HST/WFC3} passbands are indicated by name and vertical strips in different background colours. The magenta line illustrates the effect of $\Delta f_{\rm SG}$ = 0.8. For comparison, the purple line depicts the effect of $\Delta Y$ = 0.04 while the black and blue lines depict the effects of $\Delta$\,[$Z$/H] = +0.3 and $\Delta$\,log\,(age/yr) = +0.2, respectively.}
\label{f:deltaspectra}
\end{figure*}

In the top panel of Figure~\ref{f:deltaspectra} we show the resulting integrated-light spectra of FG stars and MP GCs. For the latter, we use a mixture of 20\% FG stars and 80\% SG stars (hereafter denoted as a SG fraction $f_{\rm SG}$ = 0.8) to simulate the situation as seen in massive, metal-rich GCs such as NGC 6388 and NGC 6441 \citep[cf.][]{Milone+18}. In the bottom panel of Figure~\ref{f:deltaspectra} we show ratio spectra to illustrate the effects of $\Delta f_{\rm SG} = 0.8$, $\Delta Y = 0.04$, $\Delta [Z/{\rm H}] = +0.30$, and $\Delta\log({\rm age/yr}) = +0.2$ as a function of wavelength along with some relevant \HST\ passbands, whose tables were downloaded from the \href{https://www.stsci.edu/hst}{HST website}. Note that relative to the \z\ filter passband, significant magnitude differences are seen for the $\Delta f_{SG} = 0.8$ ratio spectrum in the \UV\ and \WashC\ passbands. Specifically, ${\rm F275W}-{\rm F850LP}$ and \UVminC\ for SG stars are \emph{bluer} than that for FG stars, while ${\rm F390W}-{\rm F850LP}$ for SG stars is \emph{redder} than that for FG stars. We emphasize that this photometric signature of an increased fraction of SG stars in these filters \emph{is unique among plausible sources of stellar population gradients and is primarily driven by abundance enhancements in N and He}. As illustrated in Figure~\ref{f:deltaspectra}, increases in age and/or metallicity that are commonly found within giant ETGs 
\citep[e.g.,][]{McDermid+15, Martin-Navarro+21} cause uniform reddening in all filters in the near-UV through optical wavelength range. On the other hand, a simple increase in the helium abundance, which is often quoted as a likely cause of the UV upturn in ETGs 
\citep[e.g.,][]{Kaviraj+07,Chung+11}, mainly causes hotter stellar atmospheres and uniformly ``bluer'' colours, especially shortward of $\sim$\,450 nm. Probing and quantifying the specific photometric signature of increases of $f_{\rm SG}$ towards the centers of ETGs, and how this varies with $S_{\rm N,\,MR}$, is the observational goal of this project.  

\section{Data and Target Galaxies} \label{s:data}
The data used in this work is a combination of archival images taken with the
\emph{HST/ACS} Wide Field Camera (WFC; filters F475W and F850LP) and new images
taken with the UVIS channel of \emph{HST/WFC3} (filters F275W and F390W, program
ID 17756, PI: P.\ Goudfrooij). In this letter, we discuss the data taken for the
first two galaxies of this program, which are NGC~1380 and NGC~4649 (see
below). The analysis of the full galaxy sample as well as a detailed description
of the data reduction will be reported in a forthcoming paper once the data for
all galaxies will have been taken and analyzed. Here we only provide a
summary. Relevant properties of the data are shown in Table~\ref{tab:data}. 

\begin{table}
 \caption{Properties of \emph{HST} data.
 \label{tab:data}}
 \pgscriptsize
 \vspace*{-1.5ex}
\begin{flushleft}
 \begin{tabular}{@{}lrlllr@{}} \hline \hline
  & & & &  \\ [-2.8ex]  
   Galaxy & PID$^{(1)}$ & Obs. Date(s) & Instrument & Filter &
   $t_{\rm exp}^{(2)}$ \\ [0.5ex] \hline
  & & & &  \\ [-2.5ex]  
NGC\,1380 & 17756 &  2025 Aug 4 & WFC3/UVIS & \UV & 5869 \\
                   & 17756 &  2025 Aug 4 & WFC3/UVIS & \WashC & 1200 \\
                   & 10217 &  2004 Sep 6-7 & ACS/WFC & \g  & 760  \\
                   & 10217 &  2004 Sep 6-7 & ACS/WFC & \z & 1130  \\
NGC\,4649 & 17756 &  2025 Apr 4-21 & WFC3/UVIS & \UV & 25480 \\
                   & 17756 &  2025 Apr 5-10 & WFC3/UVIS & \WashC &  1200  \\ 
                   &  9401 &  2003 Jun 17 & ACS/WFC & \g  & 750  \\
                   &  9401 &  2003 Jun 17 & ACS/WFC & \z & 1120 \\ [0.5ex] \hline
~~ \\ [-2.2ex]
\multicolumn{6}{l}{{\it Notes:} (1): Program ID. (2): Total exposure time in seconds.}
\end{tabular}
\end{flushleft}
\end{table}

All individual images taken with a given filter are astrometrically aligned to the {\it GAIA\,DR3} reference frame and combined using the
\href{https://drizzlepac.readthedocs.io/en/latest/}{{\tt DrizzlePac} package} \citep{Fruchter+10,Hoffman+21}. Background levels of the individual images are matched using clipped medians in a region near the edge of the field of view furthest away from the galaxy center, with the same world coordinate footprint across all individual images. For each filter, the background count rates are
adjusted to the image with the lowest measured background count rate prior to
image combination using {\it AstroDrizzle}. 

Surface photometry is carried out by fitting elliptical isophotes to the combined images using the method described by \citet{jedr87} as implemented in the {\tt photutils.isophote} Python package \citep{Bradley+25}. True sky background levels for the combined images are estimated following \citet{Goudfrooij+94a}, using fits of different model functions to the outer radial intensity profiles of the galaxies. Photometric zeropoints are derived from header keywords following notebooks on the \href{https://www.stsci.edu/hst/instrumentation/wfc3/data-analysis/photometric-calibration/uvis-photometric-calibration}{WFC3 web site}.   

NGC~1380 and NGC~4649 represent good examples of ETGs near the low and high ends of both \SnMR\ and UV upturn strength, respectively: NGC~1380 has \SnMR\ = 0.38 $\pm$ 0.20 \citep{Liu+19} and (\FUVmV)$_{\rm AB,\,0}$ = 6.33 $\pm$ 0.12 within $R_{\rm e}/2$ (see Appendix), while NGC~4649 has \SnMR\ = 2.03 $\pm$ 0.47 \citep{Peng+06} and (\FUVmV)$_{\rm AB,\,0}$ = 5.51 $\pm$ 0.07 within $R_{\rm e}/2$ \citep{Goudfrooij18a}. Both galaxies are at similar distances according to \href{http://ned.ipac.caltech.edu}{NED}: 18.37 $\pm$ 0.62 Mpc for NGC~1380  and 16.70 $\pm$ 0.45 Mpc for NGC~4649, thus minimizing differential aperture effects. 

\begin{figure*}
    \includegraphics[width=8.cm]{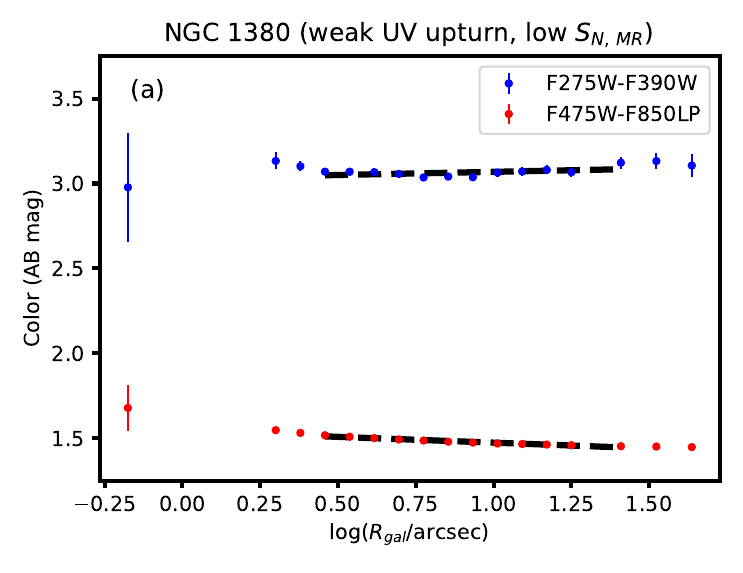}
    \hspace*{0.1mm}
    \includegraphics[width=8.cm]{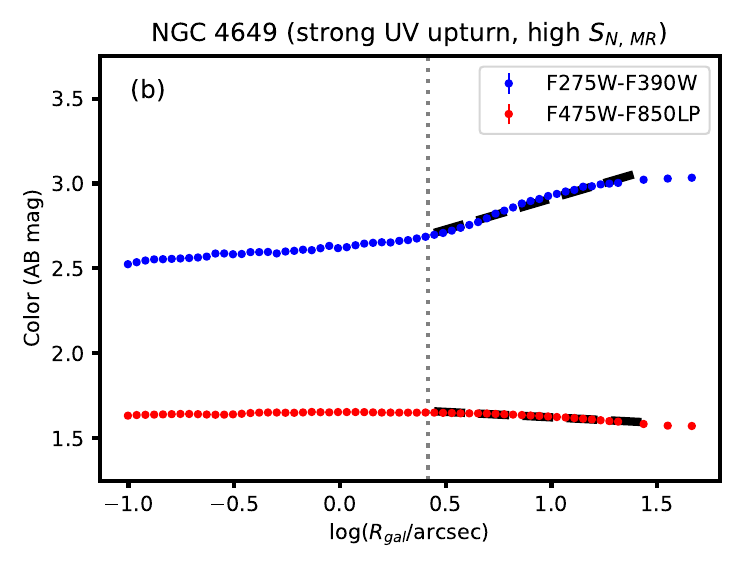}
\caption{\emph{Panel (a)}: \UVminC\ and \gminz\ colours versus log($R_{\rm gal}$) for NGC 1380. See legend for the meaning of each symbol. The black dashed lines represent linear least-square fits to colour versus log($R_{\rm gal}$). \emph{Panel (b)}: Same as Panel (a), but now for NGC 4649. The vertical dashed line indicates the radial extent of the core as measured by \citet{Ferrarese+06}. 
}
\label{f:colorgradients}
\end{figure*}

\section{Results and Discussion}   \label{s:results}
To probe possible increases in $f_{\rm SG}$ towards the centers of ETGs, we utilize radial gradients in the \UVminC\ and \gminz\ colours in an effort to disentangle the effect of variations in $f_{\rm SG}$ from gradients in metallicity, age, and/or simple helium abundance (see Sect.\ \ref{s:method} above). Previous studies of colour gradients in ETGs were mainly performed in the optical and near-IR, typically finding them to be smooth over wide intervals in galactocentric radius ($R_{\rm gal}$), becoming bluer with increasing $R_{\rm gal}$ \citep[e.g.,][]{Franx+89,Peletier+90b,Goudfrooij+94a}, while the gradients are typically flatter within the ``cores'' that are found in the innermost regions of many luminous ETGs \citep[e.g.,][]{Ferrarese+06}, possibly due to strong orbit mixing during three-body interactions between core stars and orbitally decaying supermassive black hole binaries that form in major "dry" mergers of galaxies \citep[e.g.,][]{Milosav+01,Hopkins+09,Thomas+14}. 

Radial gradients in \UVminC\ and \gminz\ for NGC 1380 and NGC 4649 are shown in Figure~\ref{f:colorgradients}. To compare the colour gradients between the two galaxies, we first exclude the innermost ``core'' region in NGC 4649 \citep[$R_{\rm core}$ = 2\farcs56,][]{Ferrarese+06}, where the colour gradients clearly flatten out, likely due to stellar orbit mixing as mentioned above \citep[see also][]{Ferrarese+06}. We also exclude the inner $\sim$\,2$''$ of NGC 1380 to avoid contamination by a inner dust disk \citep[see][]{Turner+12}. The resulting linear least-square fits to the colour gradients are shown in Figure~\ref{f:colorgradients} as dashed lines. Note that \gminz\ decreases with increasing radius for both galaxies, consistent both with other studies \citep{Ferrarese+06,Liu+11}, and with slowly decreasing [$Z$/H] and/or age with increasing radius (cf.\ Figure~\ref{f:deltaspectra}). However, the situation is very different for the \UVminC\ gradients, where \UVminC\ becomes \emph{redder} with increasing radius, which is \emph{contrary to} the expectation for simple decreases in [$Z$/H] and/or age with increasing radius, while it is consistent with an increase in $f_{\rm SG}$ towards the centers of ETGs as discussed in Sect.\ \ref{s:method}. Furthermore, the gradient in \UVminC\ is \emph{significantly} steeper for NGC 4649 than for NGC 1380 ($0.365 \pm 0.012$ vs.\ $0.051 \pm 0.026$ per dex in $R_{\rm gal}$, respectively), while this is \emph{not} the case for the gradient in \gminz\ ($-0.069 \pm 0.004$ vs.\ $-0.064 \pm 0.004$ per dex in $R_{\rm gal}$). This strongly suggests that the increase in $f_{\rm SG}$ towards the galaxy center is stronger in NGC 4649, the galaxy with a strong UV upturn and higher $S_{\it N,\,MR}$, than for NGC 1380 which has a weak UV upturn and lower $S_{\it N,\,MR}$, thus supporting the MP scenario. 

\begin{figure}
    \includegraphics[width=8.3cm]{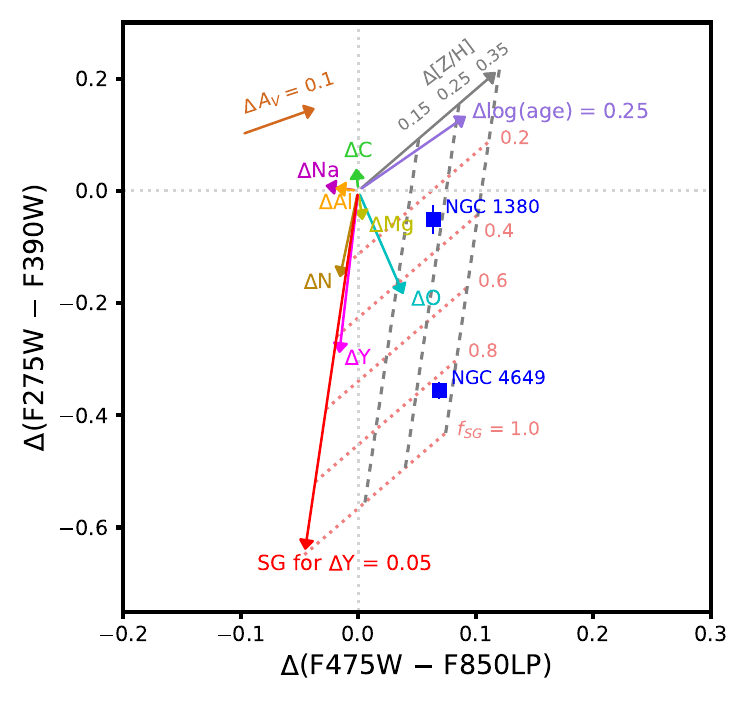}
  \caption{Radial colour gradients d(\gminz)/dlog\,($R_{\rm gal}$) and d(\UVminC)/dlog\,($R_{\rm gal}$) for NGC 1380 and NGC 4649 (blue squares) compared to SSP model predictions. Grey and purple arrows indicate the effects of increases in [$Z$/H] and log(age/yr) of 0.35 and 0.25 dex, respectively. The other arrows (except the red one) indicate the effects of the changes in individual element abundances between FG and SG stars, as mentioned in Sect.\ \ref{s:method}. Elements are indicated next to the arrowheads. The red arrow indicates the colour gradient change due to the composite effect of the individual element abundance changes mentioned above, for $\Delta Y$ = 0.05. Under the assumption of $f_{\rm SG}$ = 1.0 for $\Delta Y$ = 0.05, dashed lines connect equal values of $\Delta [Z/{\rm H}]$ for 0.00 $\leq \Delta Y \leq 0.05$, while dotted lines connect equal values of $f_{\rm SG}$ for 0.00 $\leq \Delta [Z/{\rm H}] \leq$ 0.35. A reddening vector for $\Delta\,A_V$ = 0.1 and $R_V$ = 3.1 is shown in the top left. 
  }
\label{f:colcolplot}
\end{figure}

To quantify this further, we compare the measured colour gradients for
NGC 1380 and NGC 4649 with predictions from our integrated-light SEDs
described in Sect.\ \ref{s:method}. In the colour-colour diagram in
Figure~\ref{f:colcolplot}, we draw arrows to indicate the effects of
various individual element abundance differences between FG and SG
stars as well as those of differences in overall metallicity and
age. It can be seen that for an assumed maximum helium enhancement of
$\Delta Y$ = 0.05 for the SG population (i.e., for $f_{\rm SG}$ =
1.0), the \UVminC\ and \gminz\ colour gradients for NGC 1380 and NGC
4649 can be explained by $\Delta f_{\rm SG} \simeq 0.30$ for the star
population in NGC 1380 and $\Delta f_{\rm SG} \simeq 0.85$ for that in
NGC 4649, in conjunction with moderate metallicity gradients ($\Delta
[Z/{\rm H}] \approx 0.25$ for NGC 1380 and $\Delta [Z/{\rm H}] \approx
0.30$ for NGC 4649). Note that while the derived \textit{absolute}
values for $\Delta f_{\rm SG}$ mentioned above depend on our
assumptions mentioned in Sect.\ \ref{s:method}, the \textit{relative}
values between galaxies are robust. 

\begin{figure}
    \includegraphics[width=8.3cm]{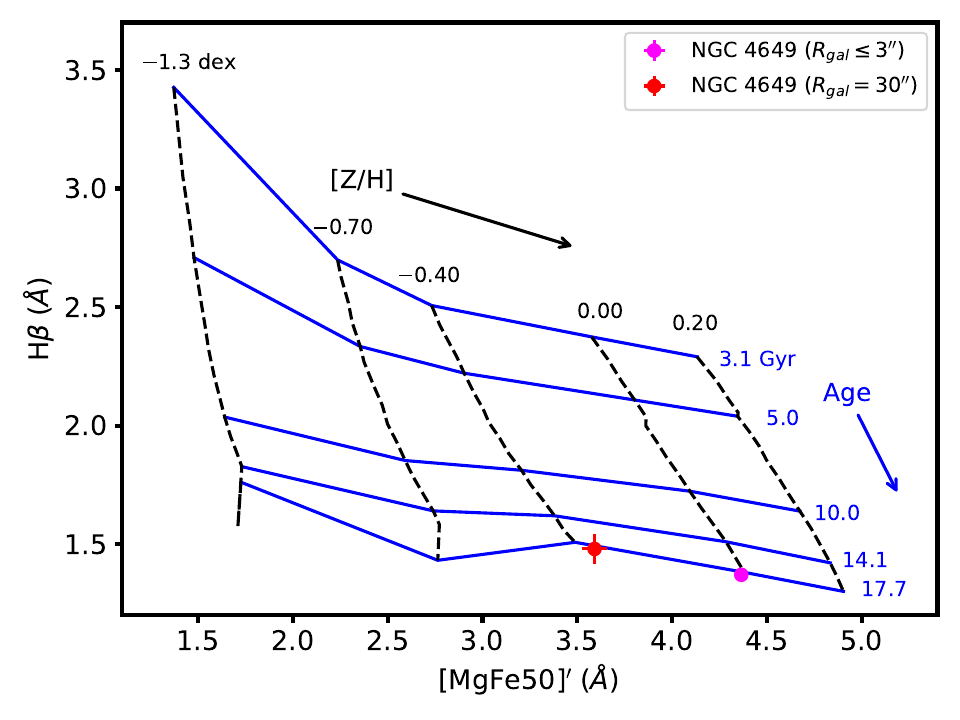}
  \caption{H$\beta$ versus [MgFe50]$'$ for NGC 4649 at $R_{\rm gal} \leq$~3$''$ and 25$'' \leq R_{\rm gal} \leq$~35$''$ from the data of \citet{McDermid+15}, overplotted with model predictions from \citet{Schiavon07} for SSPs with constant ages and metallicities (see legend). }
\label{f:schiavonplot}
\end{figure}

Note that the observed colour gradients are explained within the MSP scenario with quite reasonable values of $\Delta Y$ and $\Delta$[$Z$/H]. To illustrate the latter with results of recent spectroscopic studies, \citet{Martin-Navarro+21} found d\,[$Z$/H]/d\,log($R_{\rm gal}) = -0.25$ for NGC 1380, which is consistent with the estimate from our analysis mentioned above. For NGC 4649, we use its 2-D data table from the analysis of SAURON data by \citet{McDermid+15} to determine values of Lick indices H$\beta$, Mgb, and Fe5015 as a function of $R_{\rm gal}$. We then calculate the index [MgFe50]$'$ = (0.69 $\times$ Mgb + Fe5015)/2 which has been established as a reliable metallicity indicator that is rather insensitive to [$\alpha$/Fe] \citep{Kuntschner+10}. Values for NGC 4649 at $R_{\rm gal}$ = 3$''$ and 30$''$ are shown in Figure~\ref{f:schiavonplot} along with predictions of the SSP models of \citet{Schiavon07}. Note that the spectral data for NGC 4649 indicate a negligible age gradient along with a metallicity gradient d[$Z$/H]/dlog($R_{\rm gal}$) $\sim -0.35$, which again is consistent with that indicated by our analysis of the \HST\ photometry (see Figure~\ref{f:colcolplot}).

We emphasize that the strong positive \UVminC\ colour gradient for NGC 4649 cannot feasibly be explained by metallicity or age gradients, even in conjunction with a simple helium enhancement (i.e., one without the accompanying other light-element abundance variations seen in MSPs within massive GCs), since that would require an \emph{average} helium enhancement of $\Delta Y \sim 0.13$ (i.e., $Y \sim 0.39$ for [$Z$/H] = 0.0) for the SG population. This $\Delta Y$ would be about three times higher than that determined for any massive GC in our Galaxy \citep[see][]{Milone+18}, which seems unlikely. Moreover, such a high $\Delta Y$ would cause negative colour gradients in \gminz\ of $\sim\,-0.05$ mag per dex in $R_{\rm gal}$ (cf.\ Figure~\ref{f:colcolplot}), which would negate the influence of the known metallicity gradients in the target galaxies (see above) and is therefore inconsistent with the data. 

Finally, we address the potential contribution of diffusely
distributed dust to the observed colour gradients in the target
galaxies \citep[see, e.g.,][]{Goudfrooij+95}: the observed negative
gradient in \UVminC\ in conjunction with the consistency between the
spectroscopic metallicity gradients and those estimated from the
\UVminC\ and \gminz\ colour gradients strongly suggests that the
effect of any diffusely distributed dust on the colour gradients is
negligible (see Figure~\ref{f:colcolplot}). 

\section{Summary and Concluding Remarks}   \label{s:concl}

In an attempt to find observational evidence for (or against) the ``MP scenario'' by \citet{Goudfrooij18a} in which the UV upturn in the cores of giant ETGs is due to second-generation (SG) stars from multiple stellar populations (MPs) formed in massive star clusters and subsequently gradually released to the field through dynamical evolution in the strong tidal field in the inner regions of giant ETGs, we analyze \emph{HST/WFC3} photometry of the first two ETGs observed as part of an ongoing {\it HST} program. In this scenario, the specific frequency of metal-rich globular clusters in ETGs (\SnMR) reflects the fraction of SG stars in the inner regions of ETGs, and hence the strength of the UV upturn. The two target ETGs, NGC\,1380 and NGC\,4649, represent good examples of ETGs near the low and high ends of both \SnMR\ and UV upturn strengths, respectively. The photometric passbands used for this study are F275W and F390W, which (when combined with archival data in F475W and F850LP) are shown to be particularly sensitive to changes in the ratio of second-- to first-generation stars in integrated-light photometry, i.e., increases in $Y$ and [N/Fe] in conjunction with decreases in [O/Fe].

While the radial colour gradients in \gminz\ are found to be mildly
\emph{negative} (i.e., decreasing with increasing galactocentric radius) in both
galaxies consistent with known negative metallicity gradients, the radial
gradients in \UVminC\ are \emph{positive}, as expected if the (centrally
concentrated) UV upturn is caused by SG stars (i.e., increases of $Y$ and
[N/Fe]). Furthermore, the amplitude of the radial gradient in \UVminC\ and the
implied fractions of SG stars are found to be significantly larger in NGC\,4649
than in NGC\,1380, consistent with their respective metal-rich GC specific
frequencies $S_{\it N,\,MR}$, as predicted by the MP scenario. Finally, we show
  that the amplitudes of the colour gradients in both galaxies are consistent
  with the average values for $\Delta\,Y$ and $\Delta$\,[X/Fe] for X $\in$ [C,
    N, O, Na, Mg, Al] found within massive Galactic GCs by \citet[][and
    references therein]{Milone+18}, in conjunction with metallicity gradients
  measured from spectroscopic data in the literature.  

With regard to the question of the stellar agent(s) responsible for
the UV upturn in ETGs, the centrally concentrated nature of the latter
\citep{Ohl+98} as well as that of the He-- and N enhancement found
here suggests that the two effects are physically connected and that
their sources have a steeper radial distribution than the general
field \citep[such as stars from dissolved massive star clusters,
  see][]{Goudfrooij18a}. As such, scenarios for the UV upturn that
rely on \emph{field} stars seem less likely to produce the observed
central concentration. 

While the results of this paper look promising in terms of the feasibility of the MP scenario on the nature of the UV upturn in ETGs, they are currently based on only two galaxies. 
Results from the full observing program will be reported on in a future paper once the data from all target ETGs will have been observed and analyzed. In addition to increasing the statistics, we will also address implications with regard to dissolution processes of star clusters of various initial masses (both during the star formation era and long-term processes) in the inner regions of ETGs.  

\section*{acknowledgments}
We thank the anonymous referee for their very thoughtful comments and
suggestions which improved the presentation of this paper. 
Support for HST program {\#}17756 was provided by NASA through a grant from the Space Telescope Science Institute, which is operated by the Association of Universities for Research in Astronomy, Inc., under NASA contract NAS5-26555. THP acknowledges support from the National Agency for Research and Development (ANID) grant CATA-Basal (FB210003).

%

\section*{Data availability}
The HST data used in this paper is available at the MAST archive at
\href{https://doi.org/10.17909/nee0-qg24}{doi.org/10.17909/nee0-qg24}. The model spectra of SSPs created for this paper are available by contacting the corresponding author.

\bibliographystyle{mnras}
\bibliography{PGrefs}

@Preamble{ " \newcommand{\noop}[1]{} " }

@ARTICLE{Puzia14,
       author = {{Puzia}, Thomas H. and {Paolillo}, Maurizio and {Goudfrooij}, Paul and {Maccarone}, Thomas J. and {Fabbiano}, Giuseppina and {Angelini}, Lorella},
        title = "{Wide-field Hubble Space Telescope Observations of the Globular Cluster System in NGC 1399}",
      journal = {\apj},
         year = 2014,
        month = may,
       volume = {786},
       number = {2},
          eid = {78},
        pages = {78},
          doi = {10.1088/0004-637X/786/2/78},
archivePrefix = {arXiv},
       eprint = {1402.6714},
 primaryClass = {astro-ph.GA},
       adsurl = {https://ui.adsabs.harvard.edu/abs/2014ApJ...786...78P}
}

@ARTICLE{Brown00,
       author = {{Brown}, Thomas M. and {Bowers}, Charles W. and {Kimble}, Randy A. and
         {Sweigart}, Allen V. and {Ferguson}, Henry C.},
        title = "{Detection and Photometry of Hot Horizontal Branch Stars in the Core of M32}",
      journal = {\apj},
         year = 2000,
        month = mar,
       volume = {532},
       number = {1},
        pages = {308-322},
          doi = {10.1086/308566},
archivePrefix = {arXiv},
       eprint = {astro-ph/9909391},
 primaryClass = {astro-ph},
       adsurl = {https://ui.adsabs.harvard.edu/abs/2000ApJ...532..308B}
}

@ARTICLE{Puzia+06,
       author = {{Puzia}, Thomas H. and {Kissler-Patig}, Markus and {Goudfrooij}, Paul},
        title = "{Extremely {\ensuremath{\alpha}}-Enriched Globular Clusters in Early-Type Galaxies:A Step toward the Dawn of Stellar Populations?}",
      journal = {\apj},
         year = 2006,
        month = sep,
       volume = {648},
       number = {1},
        pages = {383-388},
          doi = {10.1086/505679},
archivePrefix = {arXiv},
       eprint = {astro-ph/0605210},
 primaryClass = {astro-ph},
       adsurl = {https://ui.adsabs.harvard.edu/abs/2006ApJ...648..383P}
}

@ARTICLE{BastianLardo18,
       author = {{Bastian}, Nate and {Lardo}, Carmela},
        title = "{Multiple Stellar Populations in Globular Clusters}",
      journal = {\araa},
         year = 2018,
        month = sep,
       volume = {56},
        pages = {83-136},
          doi = {10.1146/annurev-astro-081817-051839},
archivePrefix = {arXiv},
       eprint = {1712.01286},
 primaryClass = {astro-ph.SR},
       adsurl = {https://ui.adsabs.harvard.edu/abs/2018ARA&A..56...83B}
}

@ARTICLE{Conroy+14, 
       author = {{Conroy}, Charlie and {Graves}, Genevieve J. and {van Dokkum}, Pieter G.},
        title = "{Early-type Galaxy Archeology: Ages, Abundance Ratios, and Effective Temperatures from Full-spectrum Fitting}",
      journal = {\apj},
         year = 2014,
        month = jan,
       volume = {780},
       number = {1},
          eid = {33},
        pages = {33},
          doi = {10.1088/0004-637X/780/1/33},
archivePrefix = {arXiv},
       eprint = {1303.6629},
 primaryClass = {astro-ph.CO},
       adsurl = {https://ui.adsabs.harvard.edu/abs/2014ApJ...780...33C}
}

@ARTICLE{Kroupa01,
   author = {{Kroupa}, P.},
    title = "{On the variation of the initial mass function}",
  journal = {\mnras},
   eprint = {astro-ph/0009005},
     year = 2001,
    month = apr,
   volume = 322,
    pages = {231-246},
      doi = {10.1046/j.1365-8711.2001.04022.x},
   adsurl = {http://adsabs.harvard.edu/abs/2001MNRAS.322..231K}
}

@ARTICLE{Piotto+07, 
       author = {{Piotto}, G. and {Bedin}, L.~R. and {Anderson}, J. and {King}, I.~R. and {Cassisi}, S. and {Milone}, A.~P. and {Villanova}, S. and {Pietrinferni}, A. and {Renzini}, A.},
        title = "{A Triple Main Sequence in the Globular Cluster NGC 2808}",
      journal = {\apjl},
         year = 2007,
        month = may,
       volume = {661},
       number = {1},
        pages = {L53-L56},
          doi = {10.1086/518503},
archivePrefix = {arXiv},
       eprint = {astro-ph/0703767},
 primaryClass = {astro-ph},
       adsurl = {https://ui.adsabs.harvard.edu/abs/2007ApJ...661L..53P}
}

@ARTICLE{Worthey+92, 
       author = {{Worthey}, Guy and {Faber}, S.~M. and {Gonzalez}, J.~J.},
        title = "{MG and Fe Absorption Features in Elliptical Galaxies}",
      journal = {\apj},
         year = 1992,
        month = oct,
       volume = {398},
        pages = {69},
          doi = {10.1086/171836},
       adsurl = {https://ui.adsabs.harvard.edu/abs/1992ApJ...398...69W}
}

@ARTICLE{Trager+00, 
       author = {{Trager}, S.~C. and {Faber}, S.~M. and {Worthey}, Guy and {Gonz{\'a}lez}, J. Jes{\'u}s},
        title = "{The Stellar Population Histories of Early-Type Galaxies. II. Controlling Parameters of the Stellar Populations}",
      journal = {\aj},
         year = 2000,
        month = jul,
       volume = {120},
       number = {1},
        pages = {165-188},
          doi = {10.1086/301442},
archivePrefix = {arXiv},
       eprint = {astro-ph/0004095},
 primaryClass = {astro-ph},
       adsurl = {https://ui.adsabs.harvard.edu/abs/2000AJ....120..165T}
}

@ARTICLE{Goudfrooij18a, 
       author = {{Goudfrooij}, Paul},
        title = "{Dissolved Massive Metal-rich Globular Clusters Can Cause the Range of UV Upturn Strengths Found among Early-type Galaxies}",
      journal = {\apj},
         year = 2018,
        month = apr,
       volume = {857},
       number = {1},
          eid = {16},
        pages = {16},
          doi = {10.3847/1538-4357/aab553},
archivePrefix = {arXiv},
       eprint = {1804.00668},
 primaryClass = {astro-ph.GA},
       adsurl = {https://ui.adsabs.harvard.edu/abs/2018ApJ...857...16G}
}

@ARTICLE{Schiavon07, 
       author = {{Schiavon}, Ricardo P.},
        title = "{Population Synthesis in the Blue. IV. Accurate Model Predictions for Lick Indices and UBV Colors in Single Stellar Populations}",
      journal = {\apjs},
         year = 2007,
        month = jul,
       volume = {171},
       number = {1},
        pages = {146-205},
          doi = {10.1086/511753},
archivePrefix = {arXiv},
       eprint = {astro-ph/0611464},
 primaryClass = {astro-ph},
       adsurl = {https://ui.adsabs.harvard.edu/abs/2007ApJS..171..146S}
}

@ARTICLE{McLaughlin99, 
       author = {{McLaughlin}, Dean E.},
        title = "{The Efficiency of Globular Cluster Formation}",
      journal = {\aj},
         year = 1999,
        month = may,
       volume = {117},
       number = {5},
        pages = {2398-2427},
          doi = {10.1086/300836},
archivePrefix = {arXiv},
       eprint = {astro-ph/9901283},
 primaryClass = {astro-ph},
       adsurl = {https://ui.adsabs.harvard.edu/abs/1999AJ....117.2398M}
}

@ARTICLE{Peng+06,
       author = {{Peng}, Eric W. and {Jord{\'a}n}, Andr{\'e}s and {C{\^o}t{\'e}}, Patrick and {Blakeslee}, John P. and {Ferrarese}, Laura and {Mei}, Simona and {West}, Michael J. and {Merritt}, David and {Milosavljevi{\'c}}, Milos and {Tonry}, John L.},
        title = "{The ACS Virgo Cluster Survey. IX. The Color Distributions of Globular Cluster Systems in Early-Type Galaxies}",
      journal = {\apj},
         year = 2006,
        month = mar,
       volume = {639},
       number = {1},
        pages = {95-119},
          doi = {10.1086/498210},
archivePrefix = {arXiv},
       eprint = {astro-ph/0509654},
 primaryClass = {astro-ph},
       adsurl = {https://ui.adsabs.harvard.edu/abs/2006ApJ...639...95P}
}

@ARTICLE{Peng+08, 
       author = {{Peng}, Eric W. and {Jord{\'a}n}, Andr{\'e}s and {C{\^o}t{\'e}}, Patrick and {Takamiya}, Marianne and {West}, Michael J. and {Blakeslee}, John P. and {Chen}, Chin-Wei and {Ferrarese}, Laura and {Mei}, Simona and {Tonry}, John L. and {West}, Andrew A.},
        title = "{The ACS Virgo Cluster Survey. XV. The Formation Efficiencies of Globular Clusters in Early-Type Galaxies: The Effects of Mass and Environment}",
      journal = {\apj},
         year = 2008,
        month = jul,
       volume = {681},
       number = {1},
        pages = {197-224},
          doi = {10.1086/587951},
archivePrefix = {arXiv},
       eprint = {0803.0330},
 primaryClass = {astro-ph},
       adsurl = {https://ui.adsabs.harvard.edu/abs/2008ApJ...681..197P}
}

@ARTICLE{Gratton+19, 
       author = {{Gratton}, Raffaele and {Bragaglia}, Angela and {Carretta}, Eugenio and {D'Orazi}, Valentina and {Lucatello}, Sara and {Sollima}, Antonio},
        title = "{What is a globular cluster? An observational perspective}",
      journal = {\aapr},
         year = 2019,
        month = nov,
       volume = {27},
       number = {1},
          eid = {8},
        pages = {8},
          doi = {10.1007/s00159-019-0119-3},
archivePrefix = {arXiv},
       eprint = {1911.02835},
 primaryClass = {astro-ph.SR},
       adsurl = {https://ui.adsabs.harvard.edu/abs/2019A&ARv..27....8G}
}

@ARTICLE{Milone+17, 
       author = {{Milone}, A.~P. and {Piotto}, G. and {Renzini}, A. and {Marino}, A.~F. and {Bedin}, L.~R. and {Vesperini}, E. and {D'Antona}, F. and {Nardiello}, D. and {Anderson}, J. and {King}, I.~R. and {Yong}, D. and {Bellini}, A. and {Aparicio}, A. and {Barbuy}, B. and {Brown}, T.~M. and {Cassisi}, S. and {Ortolani}, S. and {Salaris}, M. and {Sarajedini}, A. and {van der Marel}, R.~P.},
        title = "{The Hubble Space Telescope UV Legacy Survey of Galactic globular clusters - IX. The Atlas of multiple stellar populations}",
      journal = {\mnras},
         year = 2017,
        month = jan,
       volume = {464},
       number = {3},
        pages = {3636-3656},
          doi = {10.1093/mnras/stw2531},
archivePrefix = {arXiv},
       eprint = {1610.00451},
 primaryClass = {astro-ph.SR},
       adsurl = {https://ui.adsabs.harvard.edu/abs/2017MNRAS.464.3636M}
}

@ARTICLE{Milone+18,
       author = {{Milone}, A.~P. and {Marino}, A.~F. and {Renzini}, A. and {D'Antona}, F. and {Anderson}, J. and {Barbuy}, B. and {Bedin}, L.~R. and {Bellini}, A. and {Brown}, T.~M. and {Cassisi}, S. and {Cordoni}, G. and {Lagioia}, E.~P. and {Nardiello}, D. and {Ortolani}, S. and {Piotto}, G. and {Sarajedini}, A. and {Tailo}, M. and {van der Marel}, R.~P. and {Vesperini}, E.},
        title = "{The Hubble Space Telescope UV legacy survey of galactic globular clusters - XVI. The helium abundance of multiple populations}",
      journal = {\mnras},
         year = 2018,
        month = dec,
       volume = {481},
       number = {4},
        pages = {5098-5122},
          doi = {10.1093/mnras/sty2573},
archivePrefix = {arXiv},
       eprint = {1809.05006},
 primaryClass = {astro-ph.SR},
       adsurl = {https://ui.adsabs.harvard.edu/abs/2018MNRAS.481.5098M}
}

@ARTICLE{Turner+12, 
       author = {{Turner}, Monica L. and {C{\^o}t{\'e}}, Patrick and {Ferrarese}, Laura and {Jord{\'a}n}, Andr{\'e}s and {Blakeslee}, John P. and {Mei}, Simona and {Peng}, Eric W. and {West}, Michael J.},
        title = "{The ACS Fornax Cluster Survey. VI. The Nuclei of Early-type Galaxies in the Fornax Cluster}",
      journal = {\apjs},
         year = 2012,
        month = nov,
       volume = {203},
       number = {1},
          eid = {5},
        pages = {5},
          doi = {10.1088/0067-0049/203/1/5},
archivePrefix = {arXiv},
       eprint = {1208.0338},
 primaryClass = {astro-ph.CO},
       adsurl = {https://ui.adsabs.harvard.edu/abs/2012ApJS..203....5T}
}

@ARTICLE{Marino+11,
       author = {{Marino}, A. and {Rampazzo}, R. and {Bianchi}, L. and {Annibali}, F. and {Bressan}, A. and {Buson}, L.~M. and {Clemens}, M.~S. and {Panuzzo}, P. and {Zeilinger}, W.~W.},
        title = "{Nearby early-type galaxies with ionized gas: the UV emission from GALEX observations}",
      journal = {\mnras},
         year = 2011,
        month = feb,
       volume = {411},
       number = {1},
        pages = {311-331},
          doi = {10.1111/j.1365-2966.2010.17684.x},
archivePrefix = {arXiv},
       eprint = {1009.1931},
 primaryClass = {astro-ph.CO},
       adsurl = {https://ui.adsabs.harvard.edu/abs/2011MNRAS.411..311M}
}

@ARTICLE{Renzini+15, 
       author = {{Renzini}, A. and {D'Antona}, F. and {Cassisi}, S. and {King}, I.~R. and {Milone}, A.~P. and {Ventura}, P. and {Anderson}, J. and {Bedin}, L.~R. and {Bellini}, A. and {Brown}, T.~M. and {Piotto}, G. and {van der Marel}, R.~P. and {Barbuy}, B. and {Dalessandro}, E. and {Hidalgo}, S. and {Marino}, A.~F. and {Ortolani}, S. and {Salaris}, M. and {Sarajedini}, A.},
        title = "{The Hubble Space Telescope UV Legacy Survey of Galactic Globular Clusters - V. Constraints on formation scenarios}",
      journal = {\mnras},
         year = 2015,
        month = dec,
       volume = {454},
       number = {4},
        pages = {4197-4207},
          doi = {10.1093/mnras/stv2268},
archivePrefix = {arXiv},
       eprint = {1510.01468},
 primaryClass = {astro-ph.GA},
       adsurl = {https://ui.adsabs.harvard.edu/abs/2015MNRAS.454.4197R}
}

@ARTICLE{Yong+05, 
       author = {{Yong}, D. and {Grundahl}, F. and {Nissen}, P.~E. and {Jensen}, H.~R. and {Lambert}, D.~L.},
        title = "{Abundances in giant stars of the globular cluster NGC 6752}",
      journal = {\aap},
         year = 2005,
        month = aug,
       volume = {438},
       number = {3},
        pages = {875-888},
          doi = {10.1051/0004-6361:20052916},
archivePrefix = {arXiv},
       eprint = {astro-ph/0504283},
 primaryClass = {astro-ph},
       adsurl = {https://ui.adsabs.harvard.edu/abs/2005A&A...438..875Y}
}

@ARTICLE{goukru13, 
       author = {{Goudfrooij}, Paul and {Kruijssen}, J.~M. Diederik},
        title = "{The Optical Colors of Giant Elliptical Galaxies and their Metal-Rich Globular Clusters Indicate a Bottom-Heavy Initial Mass Function}",
      journal = {\apj},
         year = 2013,
        month = jan,
       volume = {762},
       number = {2},
          eid = {107},
        pages = {107},
          doi = {10.1088/0004-637X/762/2/107},
archivePrefix = {arXiv},
       eprint = {1211.5131},
 primaryClass = {astro-ph.CO},
       adsurl = {https://ui.adsabs.harvard.edu/abs/2013ApJ...762..107G}
}

@ARTICLE{Kurucz05,
       author = {{Kurucz}, Robert L.},
        title = "{ATLAS12, SYNTHE, ATLAS9, WIDTH9, et cetera}",
      journal = {Memorie della Societa Astronomica Italiana Supplementi},
         year = 2005,
        month = jan,
       volume = {8},
        pages = {14},
       adsurl = {https://ui.adsabs.harvard.edu/abs/2005MSAIS...8...14K}
}

@ARTICLE{Castelli05,
       author = {{Castelli}, F.},
        title = "{ATLAS12: how to use it}",
      journal = {Memorie della Societa Astronomica Italiana Supplementi},
         year = 2005,
        month = jan,
       volume = {8},
        pages = {25},
       adsurl = {https://ui.adsabs.harvard.edu/abs/2005MSAIS...8...25C}
}

@INPROCEEDINGS{Sbordone+07,
       author = {{Sbordone}, Luca and {Bonifacio}, Piercarlo and {Castelli}, Fiorella},
        title = "{ATLAS 9 and ATLAS 12 under GNU-Linux}",
    booktitle = {Convection in Astrophysics},
         year = 2007,
       editor = {{Kupka}, Friedrich and {Roxburgh}, Ian and {Chan}, Kwing Lam},
       series = {IAU Symposium},
       volume = {239},
        month = may,
        pages = {71-73},
          doi = {10.1017/S1743921307000142},
       adsurl = {https://ui.adsabs.harvard.edu/abs/2007IAUS..239...71S}
}

@ARTICLE{Sbordone+11,
       author = {{Sbordone}, L. and {Salaris}, M. and {Weiss}, A. and {Cassisi}, S.},
        title = "{Photometric signatures of multiple stellar populations in Galactic globular clusters}",
      journal = {\aap},
         year = 2011,
        month = oct,
       volume = {534},
          eid = {A9},
        pages = {A9},
          doi = {10.1051/0004-6361/201116714},
archivePrefix = {arXiv},
       eprint = {1103.5863},
 primaryClass = {astro-ph.SR},
       adsurl = {https://ui.adsabs.harvard.edu/abs/2011A&A...534A...9S}
}

@ARTICLE{carbra18,
       author = {{Carretta}, Eugenio and {Bragaglia}, Angela},
        title = "{Observing multiple populations in globular clusters with the ESO archive: NGC 6388 reloaded}",
      journal = {\aap},
         year = 2018,
        month = jun, 
       volume = {614},
          eid = {A109},
        pages = {A109},
          doi = {10.1051/0004-6361/201832660},
archivePrefix = {arXiv},
       eprint = {1802.06787},
 primaryClass = {astro-ph.SR},
       adsurl = {https://ui.adsabs.harvard.edu/abs/2018A&A...614A.109C}
}

@ARTICLE{Carretta+09,
       author = {{Carretta}, E. and {Bragaglia}, A. and {Gratton}, R.~G. and {Lucatello}, S. and {Catanzaro}, G. and {Leone}, F. and {Bellazzini}, M. and {Claudi}, R. and {D'Orazi}, V. and {Momany}, Y. and {Ortolani}, S. and {Pancino}, E. and {Piotto}, G. and {Recio-Blanco}, A. and {Sabbi}, E.},
        title = "{Na-O anticorrelation and HB. VII. The chemical composition of first and second-generation stars in 15 globular clusters from GIRAFFE spectra}",
      journal = {\aap},
         year = 2009,
        month = oct,
       volume = {505},
       number = {1},
        pages = {117-138},
          doi = {10.1051/0004-6361/200912096},
archivePrefix = {arXiv},
       eprint = {0909.2938},
 primaryClass = {astro-ph.GA},
       adsurl = {https://ui.adsabs.harvard.edu/abs/2009A&A...505..117C}
}

@ARTICLE{Carretta+10, 
       author = {{Carretta}, E. and {Bragaglia}, A. and {Gratton}, R.~G. and {Recio-Blanco}, A. and {Lucatello}, S. and {D'Orazi}, V. and {Cassisi}, S.},
        title = "{Properties of stellar generations in globular clusters and relations with global parameters}",
      journal = {\aap},
         year = 2010,
        month = jun,
       volume = {516},
          eid = {A55},
        pages = {A55},
          doi = {10.1051/0004-6361/200913451},
archivePrefix = {arXiv},
       eprint = {1003.1723},
 primaryClass = {astro-ph.GA},
       adsurl = {https://ui.adsabs.harvard.edu/abs/2010A&A...516A..55C}
}

@ARTICLE{Cannon+98,
       author = {{Cannon}, R.~D. and {Croke}, B.~F.~W. and {Bell}, R.~A. and {Hesser}, J.~E. and {Stathakis}, R.~A.},
        title = "{Carbon and nitrogen abundance variations on the main sequence of 47 Tucanae}",
      journal = {\mnras},
         year = 1998,
        month = aug,
       volume = {298},
       number = {2},
        pages = {601-624},
          doi = {10.1046/j.1365-8711.1998.01671.x},
       adsurl = {https://ui.adsabs.harvard.edu/abs/1998MNRAS.298..601C}
}

@ARTICLE{Bertelli+08,
       author = {{Bertelli}, G. and {Girardi}, L. and {Marigo}, P. and {Nasi}, E.},
        title = "{Scaled solar tracks and isochrones in a large region of the Z-Y plane. I. From the ZAMS to the TP-AGB end for 0.15-2.5 \{M\}$_{{\ensuremath{\odot}}}$ stars}",
      journal = {\aap},
         year = 2008,
        month = jun,
       volume = {484},
       number = {3},
        pages = {815-830},
          doi = {10.1051/0004-6361:20079165},
archivePrefix = {arXiv},
       eprint = {0803.1460},
 primaryClass = {astro-ph},
       adsurl = {https://ui.adsabs.harvard.edu/abs/2008A&A...484..815B}
}

@ARTICLE{Gratton+06, 
       author = {{Gratton}, R.~G. and {Lucatello}, S. and {Bragaglia}, A. and {Carretta}, E. and {Momany}, Y. and {Pancino}, E. and {Valenti}, E.},
        title = "{Na-O anticorrelation and HB. III. The abundances of <ASTROBJ>NGC 6441</ASTROBJ> from FLAMES-UVES spectra}",
      journal = {\aap},
         year = 2006,
        month = aug,
       volume = {455},
       number = {1},
        pages = {271-281},
          doi = {10.1051/0004-6361:20064957},
archivePrefix = {arXiv},
       eprint = {astro-ph/0603858},
 primaryClass = {astro-ph},
       adsurl = {https://ui.adsabs.harvard.edu/abs/2006A&A...455..271G}
}

@INPROCEEDINGS{Fruchter+10, 
       author = {{Fruchter}, A.~S. and {Hack}, W. and {Dencheva}, N. and {Dr\"ottboom}, M. and {Greenfield}, P.},
        title = "{BetaDrizzle: A Redesign of the MultiDrizzle Package}",
    booktitle = {2010 Space Telescope Science Institute Calibration Workshop},
         year = 2010,
        month = jul,
        pages = {382-387},
       adsurl = {https://ui.adsabs.harvard.edu/abs/2010bdrz.conf..382F}
}

@INPROCEEDINGS{Hoffman+21,
       author = {{Hoffmann}, S.~L. and {Mack}, J. and {Avila}, R. and {Martlin}, C. and {Cohen}, Y. and {Bajaj}, V.},
        title = "{New Drizzlepac Handbook Version 2.0 Released In Hdox: Updated Documentation For HST Image Analysis}",
    booktitle = {American Astronomical Society Meeting Abstracts},
         year = 2021,
       series = {American Astronomical Society Meeting Abstracts},
       volume = {53},
        month = jun,
          eid = {216.02},
        pages = {216.02},
       adsurl = {https://ui.adsabs.harvard.edu/abs/2021AAS...23821602H}
}

@software{Bradley+25, 
  author       = {Larry Bradley and
                  Brigitta Sip{\H o}cz and
                  Thomas Robitaille and
                  Erik Tollerud and
                  Z\`e Vin{\'{\i}}cius and
                  Christoph Deil and
                  Kyle Barbary and
                  Tom J Wilson and
                  Ivo Busko and
                  Axel Donath and
                  Hans Moritz G{\"u}nther and
                  Mihai Cara and
                  P. L. Lim and
                  Sebastian Me{\ss}linger and
                  Zach Burnett and
                  Simon Conseil and
                  Michael Droettboom and
                  Azalee Bostroem and
                  E. M. Bray and
                  Lars Andersen Bratholm and
                  William Jamieson and
                  Adam Ginsburg and
                  Geert Barentsen and
                  Matt Craig and
                  Sergio Pascual and
                  Shivangee Rathi and
                  Marshall Perrin and
                  Brett M. Morris},
  title        = {astropy/photutils: 2.2.0},
  month        = feb,
  year         = 2025,
  publisher    = {Zenodo},
  version      = {2.2.0},
  doi          = {10.5281/zenodo.14889440},
  url          = {https://doi.org/10.5281/zenodo.14889440},
  swhid        = {swh:1:dir:11159107f27a28985192ed1118b1f2055709d093
                   ;origin=https://doi.org/10.5281/zenodo.596036;visi
                   t=swh:1:snp:ae8c4a55d349d43e53cfe9ce92e678fcfe840f
                   3b;anchor=swh:1:rel:0117f67e8888adcdfc85308287dd9c
                   854b466389;path=astropy-photutils-ffb96c5
                  },
}

@ARTICLE{jedr87, 
       author = {{Jedrzejewski}, Robert I.},
        title = "{CCD surface photometry of elliptical galaxies - I. Observations, reduction and results.}",
      journal = {\mnras},
         year = 1987,
        month = jun,
       volume = {226},
        pages = {747-768},
          doi = {10.1093/mnras/226.4.747},
       adsurl = {https://ui.adsabs.harvard.edu/abs/1987MNRAS.226..747J}
}

@ARTICLE{Goudfrooij+94a,
       author = {{Goudfrooij}, P. and {Hansen}, L. and {Jorgensen}, H.~E. and {N{\o}rgaard-Nielsen}, H.~U. and {de Jong}, T. and {van den Hoek}, L.~B.},
        title = "{Interstellar matter in Shapley-Ames elliptical galaxies. I. Multicolour CCD surface photometry}",
      journal = {\aaps},
         year = 1994,
        month = apr,
       volume = {104},
        pages = {179-231},
       adsurl = {https://ui.adsabs.harvard.edu/abs/1994A&AS..104..179G}
}

@ARTICLE{Goudfrooij+95, 
       author = {{Goudfrooij}, P. and {de Jong}, T.},
        title = "{Interstellar matter in Shapley-Ames elliptical galaxies. IV. A diffusely distributed component of dust and its effect on colour gradients.}",
      journal = {\aap},
         year = 1995,
        month = jun,
       volume = {298},
        pages = {784},
          doi = {10.48550/arXiv.astro-ph/9504011},
archivePrefix = {arXiv},
       eprint = {astro-ph/9504011},
 primaryClass = {astro-ph},
       adsurl = {https://ui.adsabs.harvard.edu/abs/1995A&A...298..784G}
}

@ARTICLE{Ferrarese+06, 
       author = {{Ferrarese}, Laura and {C{\^o}t{\'e}}, Patrick and {Jord{\'a}n}, Andr{\'e}s and {Peng}, Eric W. and {Blakeslee}, John P. and {Piatek}, Slawomir and {Mei}, Simona and {Merritt}, David and {Milosavljevi{\'c}}, Milo{\v{s}} and {Tonry}, John L. and {West}, Michael J.},
        title = "{The ACS Virgo Cluster Survey. VI. Isophotal Analysis and the Structure of Early-Type Galaxies}",
      journal = {\apjs},
         year = 2006,
        month = jun,
       volume = {164},
       number = {2},
        pages = {334-434},
          doi = {10.1086/501350},
archivePrefix = {arXiv},
       eprint = {astro-ph/0602297},
 primaryClass = {astro-ph},
       adsurl = {https://ui.adsabs.harvard.edu/abs/2006ApJS..164..334F}
}

@ARTICLE{Franx+89, 
       author = {{Franx}, Marijn and {Illingworth}, Garth and {Heckman}, Timothy},
        title = "{Multicolor Surface Photometry of 17 Ellipticals}",
      journal = {\aj},
         year = 1989,
        month = aug,
       volume = {98},
        pages = {538},
          doi = {10.1086/115157},
       adsurl = {https://ui.adsabs.harvard.edu/abs/1989AJ.....98..538F}
}

@ARTICLE{Peletier+90b, 
       author = {{Peletier}, R.~F. and {Valentijn}, E.~A. and {Jameson}, R.~F.},
        title = "{Near-infrared photometry of bright elliptical galaxies.}",
      journal = {\aap},
         year = 1990,
        month = jul,
       volume = {233},
        pages = {62},
       adsurl = {https://ui.adsabs.harvard.edu/abs/1990A&A...233...62P}
}

@ARTICLE{Milosav+01, 
       author = {{Milosavljevi{\'c}}, Milo{\v{s}} and {Merritt}, David},
        title = "{Formation of Galactic Nuclei}",
      journal = {\apj},
         year = 2001,
        month = dec,
       volume = {563},
       number = {1},
        pages = {34-62},
          doi = {10.1086/323830},
archivePrefix = {arXiv},
       eprint = {astro-ph/0103350},
 primaryClass = {astro-ph},
       adsurl = {https://ui.adsabs.harvard.edu/abs/2001ApJ...563...34M}
}

@ARTICLE{Thomas+14, 
       author = {{Thomas}, J. and {Saglia}, R.~P. and {Bender}, R. and {Erwin}, P. and {Fabricius}, M.},
        title = "{The Dynamical Fingerprint of Core Scouring in Massive Elliptical Galaxies}",
      journal = {\apj},
         year = 2014,
        month = feb,
       volume = {782},
       number = {1},
          eid = {39},
        pages = {39},
          doi = {10.1088/0004-637X/782/1/39},
archivePrefix = {arXiv},
       eprint = {1311.3783},
 primaryClass = {astro-ph.GA},
       adsurl = {https://ui.adsabs.harvard.edu/abs/2014ApJ...782...39T}
}

@ARTICLE{Hopkins+09, 
       author = {{Hopkins}, Philip F. and {Lauer}, Tod R. and {Cox}, Thomas J. and {Hernquist}, Lars and {Kormendy}, John},
        title = "{Dissipation and Extra Light in Galactic Nuclei. III. ``Core'' Ellipticals and ``Missing'' Light}",
      journal = {\apjs},
         year = 2009,
        month = apr,
       volume = {181},
       number = {2},
        pages = {486-532},
          doi = {10.1088/0067-0049/181/2/486},
archivePrefix = {arXiv},
       eprint = {0806.2325},
 primaryClass = {astro-ph},
       adsurl = {https://ui.adsabs.harvard.edu/abs/2009ApJS..181..486H}
}

@ARTICLE{Liu+11, 
       author = {{Liu}, Chengze and {Peng}, Eric W. and {Jord{\'a}n}, Andr{\'e}s and {Ferrarese}, Laura and {Blakeslee}, John P. and {C{\^o}t{\'e}}, Patrick and {Mei}, Simona},
        title = "{The ACS Fornax Cluster Survey. X. Color Gradients of Globular Cluster Systems in Early-type Galaxies}",
      journal = {\apj},
         year = 2011,
        month = feb,
       volume = {728},
       number = {2},
          eid = {116},
        pages = {116},
          doi = {10.1088/0004-637X/728/2/116},
archivePrefix = {arXiv},
       eprint = {1012.2634},
 primaryClass = {astro-ph.CO},
       adsurl = {https://ui.adsabs.harvard.edu/abs/2011ApJ...728..116L}
}

@ARTICLE{Liu+19, 
       author = {{Liu}, Yiqing and {Peng}, Eric W. and {Jord{\'a}n}, Andr{\'e}s and {Blakeslee}, John P. and {C{\^o}t{\'e}}, Patrick and {Ferrarese}, Laura and {Puzia}, Thomas H.},
        title = "{The ACS Fornax Cluster Survey. III. Globular Cluster Specific Frequencies of Early-type Galaxies}",
      journal = {\apj},
         year = 2019,
        month = apr,
       volume = {875},
       number = {2},
          eid = {156},
        pages = {156},
          doi = {10.3847/1538-4357/ab12d9},
archivePrefix = {arXiv},
       eprint = {1904.06909},
 primaryClass = {astro-ph.GA},
       adsurl = {https://ui.adsabs.harvard.edu/abs/2019ApJ...875..156L}
}

@ARTICLE{Martin-Navarro+21,
       author = {{Mart{\'\i}n-Navarro}, I. and {Pinna}, F. and {Coccato}, L. and {Falc{\'o}n-Barroso}, J. and {van de Ven}, G. and {Lyubenova}, M. and {Corsini}, E.~M. and {Fahrion}, K. and {Gadotti}, D.~A. and {Iodice}, E. and {McDermid}, R.~M. and {Poci}, A. and {Sarzi}, M. and {Spriggs}, T.~W. and {Viaene}, S. and {de Zeeuw}, P.~T. and {Zhu}, L.},
        title = "{Fornax 3D project: Assessing the diversity of IMF and stellar population maps within the Fornax Cluster}",
      journal = {\aap},
         year = 2021,
        month = oct,
       volume = {654},
          eid = {A59},
        pages = {A59},
          doi = {10.1051/0004-6361/202141348},
archivePrefix = {arXiv},
       eprint = {2107.14243},
 primaryClass = {astro-ph.GA},
       adsurl = {https://ui.adsabs.harvard.edu/abs/2021A&A...654A..59M}
}

@ARTICLE{McDermid+15, 
       author = {{McDermid}, Richard M. and {Alatalo}, Katherine and {Blitz}, Leo and {Bournaud}, Fr{\'e}d{\'e}ric and {Bureau}, Martin and {Cappellari}, Michele and {Crocker}, Alison F. and {Davies}, Roger L. and {Davis}, Timothy A. and {de Zeeuw}, P.~T. and {Duc}, Pierre-Alain and {Emsellem}, Eric and {Khochfar}, Sadegh and {Krajnovi{\'c}}, Davor and {Kuntschner}, Harald and {Morganti}, Raffaella and {Naab}, Thorsten and {Oosterloo}, Tom and {Sarzi}, Marc and {Scott}, Nicholas and {Serra}, Paolo and {Weijmans}, Anne-Marie and {Young}, Lisa M.},
        title = "{The ATLAS$^{3D}$ Project - XXX. Star formation histories and stellar population scaling relations of early-type galaxies}",
      journal = {\mnras},
         year = 2015,
        month = apr,
       volume = {448},
       number = {4},
        pages = {3484-3513},
          doi = {10.1093/mnras/stv105},
archivePrefix = {arXiv},
       eprint = {1501.03723},
 primaryClass = {astro-ph.GA},
       adsurl = {https://ui.adsabs.harvard.edu/abs/2015MNRAS.448.3484M}
}

@ARTICLE{Thomas+05, 
       author = {{Thomas}, Daniel and {Maraston}, Claudia and {Bender}, Ralf and {Mendes de Oliveira}, Claudia},
        title = "{The Epochs of Early-Type Galaxy Formation as a Function of Environment}",
      journal = {\apj},
         year = 2005,
        month = mar,
       volume = {621},
       number = {2},
        pages = {673-694},
          doi = {10.1086/426932},
archivePrefix = {arXiv},
       eprint = {astro-ph/0410209},
 primaryClass = {astro-ph},
       adsurl = {https://ui.adsabs.harvard.edu/abs/2005ApJ...621..673T}
}

@ARTICLE{Kuntschner+10,
       author = {{Kuntschner}, Harald and {Emsellem}, Eric and {Bacon}, Roland and {Cappellari}, Michele and {Davies}, Roger L. and {de Zeeuw}, P. Tim and {Falc{\'o}n-Barroso}, Jes{\'u}s and {Krajnovi{\'c}}, Davor and {McDermid}, Richard M. and {Peletier}, Reynier F. and {Sarzi}, Marc and {Shapiro}, Kristen L. and {van den Bosch}, Remco C.~E. and {van de Ven}, Glenn},
        title = "{The SAURON project - XVII. Stellar population analysis of the absorption line strength maps of 48 early-type galaxies}",
      journal = {\mnras},
         year = 2010,
        month = oct,
       volume = {408},
       number = {1},
        pages = {97-132},
          doi = {10.1111/j.1365-2966.2010.17161.x},
archivePrefix = {arXiv},
       eprint = {1006.1574},
 primaryClass = {astro-ph.GA},
       adsurl = {https://ui.adsabs.harvard.edu/abs/2010MNRAS.408...97K}
}

@ARTICLE{Kaviraj+07, 
       author = {{Kaviraj}, S. and {Sohn}, S.~T. and {O'Connell}, R.~W. and {Yoon}, S.-J. and {Lee}, Y.~W. and {Yi}, S.~K.},
        title = "{UV bright globular clusters in M87: more evidence for super-He-rich stellar populations?}",
      journal = {\mnras},
         year = 2007,
        month = may,
       volume = {377},
       number = {3},
        pages = {987-996},
          doi = {10.1111/j.1365-2966.2007.11712.x},
archivePrefix = {arXiv},
       eprint = {astro-ph/0703198},
 primaryClass = {astro-ph},
       adsurl = {https://ui.adsabs.harvard.edu/abs/2007MNRAS.377..987K}
}

@ARTICLE{Dorman+95, 
       author = {{Dorman}, Ben and {O'Connell}, Robert W. and {Rood}, Robert T.},
        title = "{Ultraviolet Radiation from Evolved Stellar Populations. II. The Ultraviolet Upturn Phenomenon in Elliptical Galaxies}",
      journal = {\apj},
         year = 1995,
        month = mar,
       volume = {442},
        pages = {105},
          doi = {10.1086/175428},
archivePrefix = {arXiv},
       eprint = {astro-ph/9405030},
 primaryClass = {astro-ph},
       adsurl = {https://ui.adsabs.harvard.edu/abs/1995ApJ...442..105D}
}

@ARTICLE{OConnell99, 
       author = {{O'Connell}, Robert W.},
        title = "{Far-Ultraviolet Radiation from Elliptical Galaxies}",
      journal = {\araa},
         year = 1999,
        month = jan,
       volume = {37},
        pages = {603-648},
          doi = {10.1146/annurev.astro.37.1.603},
archivePrefix = {arXiv},
       eprint = {astro-ph/9906068},
 primaryClass = {astro-ph},
       adsurl = {https://ui.adsabs.harvard.edu/abs/1999ARA&A..37..603O}
}

@ARTICLE{Ohl+98,
       author = {{Ohl}, R.~G. and {O'Connell}, R.~W. and {Bohlin}, R.~C. and {Collins}, N.~R. and {Dorman}, B. and {Fanelli}, M.~N. and {Neff}, S.~G. and {Roberts}, M.~S. and {Smith}, A.~M. and {Stecher}, T.~P.},
        title = "{Far-ultraviolet Color Gradients in Early-Type Galaxies}",
      journal = {\apjl},
         year = 1998,
        month = sep,
       volume = {505},
       number = {1},
        pages = {L11-L14},
          doi = {10.1086/311605},
archivePrefix = {arXiv},
       eprint = {astro-ph/9808265},
 primaryClass = {astro-ph},
       adsurl = {https://ui.adsabs.harvard.edu/abs/1998ApJ...505L..11O}
}

@ARTICLE{Carter+11, 
       author = {{Carter}, David and {Pass}, Sally and {Kennedy}, Joseph and {Karick}, Arna M. and {Smith}, Russell J.},
        title = "{The spatial distribution and origin of the FUV excess in early-type galaxies}",
      journal = {\mnras},
         year = 2011,
        month = jul,
       volume = {414},
       number = {4},
        pages = {3410-3423},
          doi = {10.1111/j.1365-2966.2011.18643.x},
archivePrefix = {arXiv},
       eprint = {1103.0743},
 primaryClass = {astro-ph.CO},
       adsurl = {https://ui.adsabs.harvard.edu/abs/2011MNRAS.414.3410C}
}

@ARTICLE{CodeWelch79, 
       author = {{Code}, A.~D. and {Welch}, G.~A.},
        title = "{Ultraviolet photometry from the Orbiting Astronomical Observatory. XXVI. Energy distributions of seven early-type galaxies and the central bulge of M31.}",
      journal = {\apj},
         year = 1979,
        month = feb,
       volume = {228},
        pages = {95-104},
          doi = {10.1086/156825},
       adsurl = {https://ui.adsabs.harvard.edu/abs/1979ApJ...228...95C}
}

\clearpage
\appendix
\section{Determination of $({\it FUV}-V)_{\rm AB,\,0}$ for NGC 1380}
To determine $({\it FUV}-V)_{\rm AB,\,0}$ within $R_{\rm e}$/2 for NGC\,1380, we use the \emph{GALEX}/FUV surface brightness measurements by \citet{Marino+11} along with their assumptions of $R_{\rm e}$ = 20\farcs3 and Galactic extinction ($R_V$ = 3.1, $E_{B-V}$ = 0.015, $A_{\rm FUV}/E_{B-V}$ = 8.376). To obtain $V$-band photometry for NGC\,1380, we use our F475W surface brightness profile (cf.\ Sect.\ \ref{s:data}), scaled to the multi-aperture $V$-band photometry from \citet*{Persson+79} after  integrating the surface brightness along the circularized radius $R_{\rm gal} = a\sqrt{1-\epsilon}$ where $a$ is the distance along the semi-major axis of the ellipse and $\epsilon$ its ellipticity. We obtain ${\it FUV}_{\rm AB,\,0}$ = 18.24 $\pm$ 0.12 and $V_0$ = 11.91 $\pm$ 0.01 within $R_{\rm e}$/2 using spline interpolation, thus resulting in $({\it FUV}-V)_{\rm AB,\,0}$ = 6.33 $\pm$ 0.12. 



\end{document}